

%
%

\font\twelverm=cmr10 scaled 1200    \font\twelvei=cmmi10 scaled 1200
\font\twelvesy=cmsy10 scaled 1200   \font\twelveex=cmex10 scaled 1200
\font\twelvebf=cmbx10 scaled 1200   \font\twelvesl=cmsl10 scaled 1200
\font\twelvett=cmtt10 scaled 1200   \font\twelveit=cmti10 scaled 1200
\font\twelvesc=cmcsc10 scaled 1200  
\skewchar\twelvei='177   \skewchar\twelvesy='60


\def\twelvepoint{\normalbaselineskip=12.4pt plus 0.1pt minus 0.1pt
  \abovedisplayskip 12.4pt plus 3pt minus 9pt
  \belowdisplayskip 12.4pt plus 3pt minus 9pt
  \abovedisplayshortskip 0pt plus 3pt
  \belowdisplayshortskip 7.2pt plus 3pt minus 4pt
  \smallskipamount=3.6pt plus1.2pt minus1.2pt
  \medskipamount=7.2pt plus2.4pt minus2.4pt
  \bigskipamount=14.4pt plus4.8pt minus4.8pt
  \def\rm{\fam0\twelverm}          \def\it{\fam\itfam\twelveit}%
  \def\sl{\fam\slfam\twelvesl}     \def\bf{\fam\bffam\twelvebf}%
  \def\mit{\fam 1}                 \def\cal{\fam 2}%
  \def\sc{\twelvesc}               \def\tt{\twelvett}
  \def\sf{\twelvesf}
  \textfont0=\twelverm   \scriptfont0=\tenrm   \scriptscriptfont0=\sevenrm
  \textfont1=\twelvei    \scriptfont1=\teni    \scriptscriptfont1=\seveni
  \textfont2=\twelvesy   \scriptfont2=\tensy   \scriptscriptfont2=\sevensy
  \textfont3=\twelveex   \scriptfont3=\twelveex  \scriptscriptfont3=\twelveex
  \textfont\itfam=\twelveit
  \textfont\slfam=\twelvesl
  \textfont\bffam=\twelvebf \scriptfont\bffam=\tenbf
  \scriptscriptfont\bffam=\sevenbf
  \normalbaselines\rm}



\def\beginlinemode{\endmode
  \begingroup\parskip=0pt \obeylines\def\\{\par}\def\endmode{\par\endgroup}}
\def\beginparmode{\endmode
  \begingroup \def\endmode{\par\endgroup}}
\let\endmode=\par
{\obeylines\gdef\
{}}
\def\singlespace{\baselineskip=\normalbaselineskip}

\def\oneandahalfspace{\baselineskip=\normalbaselineskip
  \multiply\baselineskip by 3 \divide\baselineskip by 2}
\def\doublespace{\baselineskip=\normalbaselineskip \multiply\baselineskip by 2}

\newcount\firstpageno
\firstpageno=2
\footline={\ifnum\pageno<\firstpageno{\hfil}\else{\hfil\twelverm\folio\hfil}\fi}
\def\toppageno{\global\footline={\hfil}\global\headline
  ={\ifnum\pageno<\firstpageno{\hfil}\else{\hfil\twelverm\folio\hfil}\fi}}
\let\rawfootnote=\footnote              
\def\footnote#1#2{{\rm\singlespace\parindent=0pt\parskip=0pt
  \rawfootnote{#1}{#2\hfill\vrule height 0pt depth 6pt width 0pt}}}
\def\raggedcenter{\leftskip=4em plus 12em \rightskip=\leftskip
  \parindent=0pt \parfillskip=0pt \spaceskip=.3333em \xspaceskip=.5em
  \pretolerance=9999 \tolerance=9999
  \hyphenpenalty=9999 \exhyphenpenalty=9999 }
\def\dateline{\rightline{\ifcase\month\or
  January\or February\or March\or April\or May\or June\or
  July\or August\or September\or October\or November\or December\fi
  \space\number\year}}
\def\received{\vskip 3pt plus 0.2fill
 \centerline{\sl (Received\space\ifcase\month\or
  January\or February\or March\or April\or May\or June\or
  July\or August\or September\or October\or November\or December\fi
  \qquad, \number\year)}}


\hsize=6.5truein
\vsize=8.5truein  
\voffset=-1.0truein
\parskip=\medskipamount
\def\\{\cr}
\twelvepoint            
\doublespace            
\overfullrule=0pt       

\def\title                      
  {\null\vskip 3pt plus 0.2fill
   \beginlinemode \doublespace \raggedcenter \bf}

\def\author                     
  {\vskip 3pt plus 0.2fill \beginlinemode
   \singlespace \raggedcenter\sc}

\def\affil                      
  {\vskip 3pt plus 0.1fill \beginlinemode
   \oneandahalfspace \raggedcenter \sl}

\def\abstract                   
  {\vskip 3pt plus 0.3fill \beginparmode
   \singlespace ABSTRACT: }

\def\endtopmatter               
  {\endpage                     
   \body}

\def\body                       
  {\beginparmode}               

\def\head#1{                    
  \goodbreak\vskip 0.5truein    
  {\immediate\write16{#1}
   \raggedcenter \uppercase{#1}\par}
   \nobreak\vskip 0.25truein\nobreak}

\def\subhead#1{                 
  \vskip 0.25truein             
  {\raggedcenter {#1} \par}
   \nobreak\vskip 0.25truein\nobreak}

\def\beginitems{
\par\medskip\bgroup\def\i##1 {\item{##1}}\def\ii##1 {\itemitem{##1}}
\leftskip=36pt\parskip=0pt}
\def\enditems{\par\egroup}

\def\beneathrel#1\under#2{\mathrel{\mathop{#2}\limits_{#1}}}

\def\refto#1{$^{#1}$}           

\def\references                 
  {\head{References}            
   \beginparmode
   \frenchspacing \parindent=0pt \leftskip=1truecm
   \parskip=8pt plus 3pt \everypar{\hangindent=\parindent}}

\gdef\refis#1{\item{#1.\ }}                     

\gdef\journal#1, #2, #3, 1#4#5#6{               
    {\sl #1~}{\bf #2}, #3 (1#4#5#6)}            

\gdef\refa#1, #2, #3, #4, 1#5#6#7.{\noindent#1, #2 {\bf #3}, #4 (1#5#6#7).\rm}

\gdef\refb#1, #2, #3, #4, 1#5#6#7.{\noindent#1 (1#5#6#7), #2 {\bf #3}, #4.\rm}

\def\pr{\journal Phys.Rev., }

\def\prl{\journal Phys.Rev.Lett., }

\def\cmp{\journal Comm.Math.Phys., }

\def\pl{\journal Phys.Lett., }

\def\annp{\journal Ann.Phys.(N.Y.), }

\def\endreferences{\body}

\def\endpage                    
  {\vfill\eject}

\def\endpaper                   
  {\endmode\vfill\supereject}

\def\ref#1{Ref.~#1}                     
\def\Ref#1{Ref.~#1}                     
\def\[#1]{[\cite{#1}]}
\def\cite#1{{#1}}
\def\(#1){(\call{#1})}
\def\call#1{{#1}}
\def\taghead#1{}
\def\frac#1#2{{#1 \over #2}}
\def\half{{\frac 12}}

\def\12{{1\over2}}

\catcode`@=11
\newcount\r@fcount \r@fcount=0
\newcount\r@fcurr
\immediate\newwrite\reffile
\newif\ifr@ffile\r@ffilefalse
\def\w@rnwrite#1{\ifr@ffile\immediate\write\reffile{#1}\fi\message{#1}}

\def\writer@f#1>>{}
\def\referencefile{
  \r@ffiletrue\immediate\openout\reffile=\jobname.ref%
  \def\writer@f##1>>{\ifr@ffile\immediate\write\reffile%
    {\noexpand\refis{##1} = \csname r@fnum##1\endcsname = %
     \expandafter\expandafter\expandafter\strip@t\expandafter%
     \meaning\csname r@ftext\csname r@fnum##1\endcsname\endcsname}\fi}%
  \def\strip@t##1>>{}}

\def\citeall#1{\xdef#1##1{#1{\noexpand\cite{##1}}}}
\def\cite#1{\each@rg\citer@nge{#1}}	

\def\each@rg#1#2{{\let\thecsname=#1\expandafter\first@rg#2,\end,}}
\def\first@rg#1,{\thecsname{#1}\apply@rg}	
\def\apply@rg#1,{\ifx\end#1\let\next=\relax
\else,\thecsname{#1}\let\next=\apply@rg\fi\next}

\def\citer@nge#1{\citedor@nge#1-\end-}	
\def\citer@ngeat#1\end-{#1}
\def\citedor@nge#1-#2-{\ifx\end#2\r@featspace#1 
  \else\citel@@p{#1}{#2}\citer@ngeat\fi}	
\def\citel@@p#1#2{\ifnum#1>#2{\errmessage{Reference range #1-#2\space is bad.}%
    \errhelp{If you cite a series of references by the notation M-N, then M and
    N must be integers, and N must be greater than or equal to M.}}\else%
 {\count0=#1\count1=#2\advance\count1 by1\relax\expandafter\r@fcite\the\count0,
  \loop\advance\count0 by1\relax
    \ifnum\count0<\count1,\expandafter\r@fcite\the\count0,%
  \repeat}\fi}

\def\r@featspace#1#2 {\r@fcite#1#2,}	
\def\r@fcite#1,{\ifuncit@d{#1}
    \newr@f{#1}%
    \expandafter\gdef\csname r@ftext\number\r@fcount\endcsname%
                     {\message{Reference #1 to be supplied.}%
                      \writer@f#1>>#1 to be supplied.\par}%
 \fi%
 \csname r@fnum#1\endcsname}
\def\ifuncit@d#1{\expandafter\ifx\csname r@fnum#1\endcsname\relax}%
\def\newr@f#1{\global\advance\r@fcount by1%
    \expandafter\xdef\csname r@fnum#1\endcsname{\number\r@fcount}}

\let\r@fis=\refis			
\def\refis#1#2#3\par{\ifuncit@d{#1}
   \newr@f{#1}%
   \w@rnwrite{Reference #1=\number\r@fcount\space is not cited up to now.}\fi%
  \expandafter\gdef\csname r@ftext\csname r@fnum#1\endcsname\endcsname%
  {\writer@f#1>>#2#3\par}}

\def\ignoreuncited{
   \def\refis##1##2##3\par{\ifuncit@d{##1}%
    \else\expandafter\gdef\csname r@ftext\csname r@fnum##1\endcsname\endcsname%
     {\writer@f##1>>##2##3\par}\fi}}

\def\r@ferr{\endreferences\errmessage{I was expecting to see
\noexpand\endreferences before now;  I have inserted it here.}}
\let\r@ferences=\references
\def\references{\r@ferences\def\endmode{\r@ferr\par\endgroup}}

\let\endr@ferences=\endreferences
\def\endreferences{\r@fcurr=0
  {\loop\ifnum\r@fcurr<\r@fcount
    \advance\r@fcurr by 1\relax\expandafter\r@fis\expandafter{\number\r@fcurr}%
    \csname r@ftext\number\r@fcurr\endcsname%
  \repeat}\gdef\r@ferr{}\endr@ferences}


\let\r@fend=\endpaper\gdef\endpaper{\ifr@ffile
\immediate\write16{Cross References written on []\jobname.REF.}\fi\r@fend}

\catcode`@=12

\citeall\refto		
\citeall\ref		%
\citeall\Ref		%


\oneandahalfspace

\def\A{{\cal A}}
\def\D{\Delta}
\def\p{\partial}
\def\la{\langle}
\def\ra{\rangle}
\def\ria{\rightarrow}

\def\s{{\sigma}}
\def\a{\alpha}
\def\b{\beta}
\def\e{\epsilon}

\def\om{{\omega}}

\def\ih{{ {i \over \hbar} }}
\def\trho{{\rho}}


\def\jjh{{j.halliwell@ic.ac.uk}}

\centerline{\bf Generalized Uncertainty Relations and Long Time Limits}
\centerline{\bf for Quantum Brownian Motion Models}

\vskip 0.3in
\author Charalambos Anastopoulos
\vskip 0.2in
\centerline {and}
\author Jonathan J. Halliwell\footnote{$^{*}$}{E-mail address:\jjh}
\affil
Theory Group
Blackett Laboratory
Imperial College
South Kensington
London SW7 2BZ
UK
\vskip 0.5in
\centerline {\rm Preprint IC 92-93/25. July, 1994}
\vskip 0.2in
\centerline {\rm Submitted to {\sl Physical Review D}}

\abstract
{We study the time evolution of the reduced density operator for a
class of quantum Brownian motion models consisting of a particle
moving in a potential $V(x)$ and coupled to an environment of
harmonic oscillators in a thermal state.
Our principle tool is the Wigner function of the reduced density
operator, and for linear systems we derive an explicit expression
for the Wigner function propagator.
We use it to derive two generalized uncertainty relations. The first
consists of a sharp lower bound on the uncertainty function,
$U = (\Delta p)^2 (\Delta q)^2 $, after evolution for time $t$ in the
presence of an environment. The second, a stronger and simpler result,
consists of a lower bound at time $t$ on the quantity,
$ \A^2 = U - C_{pq}^2 $, where
$ C_{pq} = \half \langle \Delta \hat p \Delta \hat q +
\Delta \hat q \Delta \hat p \rangle$. ($\A$ is essentially
the area enclosed by the $1-\sigma$ contour of the Wigner function).
In both cases the minimizing
initial state is a correlated coherent state (a non-minimal
Gaussian pure state), and in the first case the lower bound is only an
envelope. These generalized uncertainty relations supply a measure
of the comparative size of quantum and thermal fluctuations.
We prove two simple inequalites, relating uncertainty to von Neumann entropy,
$-{\rm Tr}(\rho \ln \rho)$, and the von Neumann entropy to linear
entropy, $ 1 - {\rm Tr} \rho^2 $.
We also prove some results on the long-time limit of the Wigner
function for arbitrary initial states.
For the harmonic oscillator the Wigner function for
all initial states becomes a Gaussian at large times (often, but not
always, a thermal state). We derive the explicit forms of the
long-time limit for the free particle (which does not in general
go to a Gaussian), and also for more general potentials in the
approximation of high temperature.
We discuss connections with previous work by Hu and Zhang and by
Paz and Zurek.
}
\endtopmatter

\head{\bf 1.Introduction}

Quantum Brownian motion models have been the subject of a number of
studies of many years [\cite{Aga,Ang,Bru,CaL,Dek,DoH,FLO,GSI,HuR,HuZ,TeS}].
Many reasons for the interest in these
models may be found: they permit the
possibility of studying in some detail the approach to equilibirum in
non-equilibrium system; they arise in studies of macroscopic
quantum effects; and they are related to the question of dissipation in
tunneling. Most recently, they have been studied in the contexts of
quantum measurement theory, decoherence, and the quantum to
classical transition. It is these contexts towards which the present
work is directed. We are, in particular, interested in the emergence
of classical behaviour in open quantum systems.

The quantum Brownian motion models belong to an important class of
non-equilibrium systems in which there is a natural separation into
a distinguished subsystem, ${\cal S}$, and the rest (the
environment). The distinguished subsystem ${\cal S}$ is often
referred to as an open quantum system. One is interested in the
behaviour of ${\cal S}$, but not in the detailed behaviour of the
environment. ${\cal S}$ is most completely describe by the reduced
density operator $\rho$ obtained by tracing out over the environment
states. One's goal is then to obtain an evolution equation for
$\rho$, which will in general be non-unitary, from which one may
calculate the probabilities of any observables referring to ${\cal
S}$ only.
In the one-dimensional QBM models studied in this paper,
${\cal S}$ consists of a
particle of mass $M$ moving in a potential $V(x)$. ${\cal S}$ is
linearly coupled to an environment, consisting of a large number of
harmonic oscillators in a thermal state at temperature $T$, and
characterized by a dissipation coefficient $\gamma$.

Given such a model, there are many interesting questions one can
then ask about it.  Under what conditions is there suppression of
interference between localized wave packets? Under what conditions
does the Brownian particle evolve approximately classically?  How
big are the fluctuations about classical predictability? Are these
fluctuations larger than the inescapable quantum fluctuations? Is
there a generalization of the uncertainty principle to include
environmentally-induced fluctuations, representing the smallest
amount of noise the system must suffer? What sort of states are most
stable in the face of these fluctuations? Does entropy/uncertainty
increase as time evolves?  Towards what sort of states does the
system evolve in the long-time limit? Does the system tend towards
thermal equilibrium in the long-time limit?

Many of these questions have been addressed before
[\cite{AnH,HuZ,JoZ,PHZ,PaZ,TeS,UnZ,ZHP}]
and some of
them will be the topic of the present paper. We will address in
particular, the question of the generalization of the uncertainty
principle, and the question of the long-time limit of arbitrary initial
states. The methods we will develop, however, will be applicable to
most of the other questions listed above.

Our starting point is the observation that all of the above
questions depend on the dynamical evolution of the system. It is
therefore of crucial importance to possess a clear picture of that
evolution. This we obtain by focusing on the evolution of the Wigner
function of the reduced density operator. In particular, the tool we
found to be of greatest use is the Wigner function propagator.
The reason why the Wigner representation is so clarifying is that,
at least for linear systems,
the unitary part of the evolution corresponds to transporting
the Wigner function along the classical phase space trajectories.
The Wigner function propagator in this case is then just a
product of delta-functions.
The same holds to leading order in an $\hbar $ expansion for
non-linear systems. It is therefore possible to cleanly separate
the unitary effects from the non-unitary effects induced by the
environment.

In the presence of an environment, the Wigner function propagator
may be calculated exactly for linear systems, and it is just a Gaussian.
It permits the
explicit calculation of all moments of $p$ and $q$ at any
time for arbitrary initial states in terms of the moments at the
initial time. For systems with more general potentials, we cannot
calculate it exactly, but some interesting
aspects of the evoluion may be
extracted using a semiclassical approximation.

We describe the features of quantum Brownian motion models in
Section II, and we calculate the Wigner function propagator for
linear systems in Section III.

In Section IV, we use the results of Section III to derive
generalized uncertainty relations for quantum Brownian motion
models. Two relations are derived. The first is a lower bound over
all initial states on the uncertainty function
$$
U =  ( \D p )^2 (\D q )^2
\eqno(1.1)
$$
at an arbitrary time $t$. This lower bound represents the least
amount of noise, both quantum and environmentally-induced,
the system must suffer after evolution for time $t$ in the
presence of an environment.
The lower bound is not the time evolution of a particular initial
state, but is actually an envelope -- the initial state achieving
the lower bound at time $t$ is different for each time. The second
relation is a lower bound on the related quantity,
$$
{\cal A}^2 = (\D p )^2 (\D q)^2 - {1 \over 4} \langle \D \hat  p \D
\hat q + \D \hat q  \D \hat p \rangle^2
\eqno(1.2)
$$
where $\D \hat q = \hat q - \la \hat q \ra $ and likewise for $\D
\hat p$.
${\cal A} $ is essentially the area enclosed by the $1-\sigma$
contour of the Wigner function. A slightly stronger version of the
usual uncertainty principle exists in terms of ${\cal A}$, taking
the form ${\cal A} \ge \hbar /2 $ [\cite{DKM}].
The time evolution of ${\cal A}$
turns out to be much simpler than that of $U$, and the lower bound
on it at any time {\it is} the time evolution of a particular
initial state, hence this result is stronger than the first one.
In both case the minimizing initial state is a {\it correlated
coherent state}. A non-minimal Gaussian pure state, of the form
$ \psi(x) = e^{- (a + ib) x^2 } $, where $a$, $b$ are real.
(This is clearly some kind of {\it squeezed} coherent state, but we
choose to follow the nomenclature of Ref.[\cite{DKM}]).
The detailed forms of the lower bounds on $U$ and ${\cal A}$
are discussed in Section V, in the case of an ohmic environment.

In Section VI we discuss the connection of these measures of
uncertainty with von Neumann entropy, $ S[\rho] = - {\rm Tr}(\rho
\ln \rho) $. We prove an inequality relating uncertainty to von
Neumann entropy, which in the regime of large uncertainty has the form,
$ U \ge {\cal A}^2 \ge \hbar^2 e^{2 S} $. We also exhibit the
connection with the linear entropy, $ S_L = 1 - {\rm Tr} \rho^2 $.

In Section VII, we consider the long-time limit of an arbitrary initial
state. For the harmonic oscillator in an ohmic environment, all
initial states go to a Gaussian Wigner function in the long-time
limit, which is not always a thermal state.
In Section VIII we show how some of our results
may be generalized to systems with more general potentials.
We also consider the case of the free particle.

We summarize and discuss our results in Section IX.

\head{\bf 2. Quantum Brownian Motion Models}

We are concerned in this paper with the class of
quantum Brownian models consisting of a particle of large mass $M$
moving in a potential $V(x)$ and linearly coupled to a
bath of harmonic oscillators. (This section is entirely review of
standard material [\cite{FeV,CaL,GSI,HPZ}].)
The total system is therefore described by the action,
$$
\eqalignno{
S_{Tot}[x(t), q_n(t)] = &\int dt \left[ \half M \dot x^2 - V(x) \right]
\cr &
+ \sum_n \int dt \left[ \half m_n \dot q_n^2 - \half m_n \om_n^2 q_n^2
- C_n q_n x \right]
&(2.1) \cr}
$$
Quantum-mechanically, the total system will be described most
completely by the density matrix, $\rho_t(x,q_n,x', q_n')$. However,
we are interested solely in the behaviour of the large particle, and
hence the relevant quantity is the reduced density matrix, obtained
by tracing over the environmental coordinates,
$$
\rho_t (x,x') = \prod_n \int dq_n \ \rho_t(x,q_n,x', q_n)
\eqno(2.2)
$$
It is convenient to introduce the (non-unitary) reduced
density matrix propagator, $J$, defined by the relation,
$$
\rho_t (x,y) = \int dx_0 dy_0 \ J(x,y,t|x_0,y_0,0) \ \rho_0 (x_0, y_0)
\eqno(2.3)
$$
Under the assumption that the initial density operator for the total
system factorizes, $\rho = \rho_{system} \otimes \rho_{bath}$, the
reduced density operator propagator is given by
the path integral expression,
$$
J(x_f,y_f,t|x_0,y_0,0) = \int {\cal D}x {\cal D}y \ \exp \left(
\ih S[x] - \ih S[y] + \ih W[x,y] \right)
\eqno(2.4)
$$
where
$$
S[x] =  \int dt \left[ \half M \dot x^2 - V(x) \right]
\eqno(2.5)
$$
and $W[x(t),y(t)]$ is the Feynman-Vernon influence functional phase,
$$
\eqalignno{
W[x(t),y(t)] = & -
\int_0^t ds \int_0^s ds' [ x(s) - y(s) ] \ \eta (s-s') \ [ x(s') + y(s') ]
\cr &
+ i \int_0^t ds \int_0^s ds' [ x(s) - y(s) ] \ \nu(s-s') \ [ x(s') - y(s') ]
&(2.6) \cr }
$$
The kernels $\eta(s)$ and $\nu(s)$ are defined by
$$
\eqalignno{
\nu(s) = & \int_0^{\infty} { d \om \over \pi} I(\om)
\coth\left( {\hbar \om \over 2 kT} \right) \cos \om s
&(2.7) \cr
\eta(s) = & { d \over ds} \gamma(s)
&(2.8) \cr }
$$
Here
$$
\gamma(s) =
\int_0^{\infty} { d \om \over \pi} { I(\om) \over \om}
\cos \om s
\eqno(2.9)
$$
and $I(\om)$ is the spectral density
$$
I(\om) = \sum_n \delta (\om - \om_n) { \pi C_n^2 \over2 m_n \om_n}
\eqno(2.10)
$$

The kernel $\nu(s)$ contributes a phase to the path integral (2.4),
effectively modifying the action of the distinguished system. It
leads to dissipation and frequency renormalization in the effective
equations of motion. The kernel $\eta(s)$ damps contributions from
differing values of $x$ and $y$. It is responsible for noise
(and also for the process of decoherence discussed elsewhere).
These two kernels are completely determined once a form for the
spectral density (2.10) has been specified. A convenient class
of choices is,
$$
I (\om) = M \gamma \om \left( { \om \over \tilde \om} \right)^{s-1}
\ \exp \left(- { \om^2 \over \Lambda^2 } \right)
\eqno(2.11)
$$
Here, $\Lambda$ is a cut-off, which will generally be taken to be
very large, sometimes infinite, and $\tilde \om$ is a frequency
scale, which may be taken to be $\Lambda$. We will
concentrate almost entirely
on the case of the ohmic environment, $s=1$, with
occassional reference to the supra-ohmic and subohmic cases,
$s>1$ and $s<1$ respectively.

In the ohmic case, (2.8) is,
$$
\gamma(s) = M \gamma { \Lambda \over 2 \pi^{\half} }
\ \exp \left( - {1 \over 4} \Lambda^2 s^2 \right)
\eqno(2.12)
$$
and thus when $\Lambda$ is very large,
$$
\gamma(s) \approx M \gamma \delta(s)
\eqno(2.13)
$$
We will work in this limit, unless otherwise stated.
It should be noted, however, that this limit sometimes leads to a
violation of positivity of the density operator at very short
timescales (of order $\Lambda^{-1}$) [\cite{Amb}].
We will say more about this later.

In the limit in which (2.13) holds, the real
part of the influence functional phase (2.6) may be written,
$$
\eqalignno{
Re W = & \int_0^t ds \ M \gamma (x-y) (\dot x - \dot y)
\cr &
- \int_0^t  ds \ M \gamma \ \delta(0) \ (x^2 -y^2)
+ M \gamma ( x(0)^2 - y(0)^2 )
&(2.14) \cr }
$$
Here, the $\delta(0)$ is understood in terms of (2.12) at $s=0$ for
large $\Lambda$. It is easily seen that the terms involving it may
be absorbed by defining a renormalized potential in the path
integral (2.4),
$$
V_R(x) = V(x) - M \gamma \delta(0) x^2
\eqno(2.15)
$$
For a harmonic oscillator of frequency $\om_0$, considered below, we
therefore define a renormalized frequency, $\om_R$, with $\om_R^2 =
\om^2_0 - 2 \gamma \delta (0) $. The terms involving the end-points
$x(0)$, $y(0)$ are clearly negligible and will be dropped.

The noise kernel (2.7) is non-local for large $\Lambda$, except in
the so-called Fokker-PLanck limit, $ kT \ >> \ \hbar \Lambda$,  in which
case one has
$$
\nu(s) = { 2 M \gamma k T \over \hbar} \ \delta(s)
\eqno(2.16)
$$

An evolution equation for $\rho$ may be derived.
Its expected most general form is,
$$
\eqalignno{
i \hbar { \partial  \trho \over \partial t} =
& - { \hbar^2 \over 2M } \left( { \partial^2 \trho \over \partial x^2 }
- { \partial^2 \trho \over \partial y^2 } \right)
+ \left[ V_R(x) - V_R(y) \right] \trho
\cr &
- i \hbar \Gamma (t) (x-y) \left( { \partial \trho \over \partial x}
- { \partial  \trho \over \partial y} \right)
- i \Gamma (t) h(t) (x-y)^2 \trho
\cr &
+ \hbar \Gamma(t) f(t) (x-y) \left( {\partial \rho \over \partial x}
+ {\partial \rho \over \partial y} \right)
&(2.17) \cr }
$$
The coefficients
$\Gamma(t)$, $f(t)$, $h(t)$, are in general
rather complicated non-local functions of time, and appear
to be known only in two particular cases.
Firstly,
explicit expressions for them may be found in
Ref.[\cite{HPZ}] for the case of linear systems.
Secondly,
in the Fokker-Planck limit, for any potential $V(x)$, one has [\cite{CaL}],
$$
\Gamma(t) = \gamma , \quad h(t) = { 2 M kT \over \hbar}, \quad f(t) = 0
\eqno(2.18)
$$

The propagator $J$ may be evaluated exactly for the case of the
simple harmonic oscillator, $ V(x) = \half M \omega^2_0 x^2 $.
We will also be interested in the free particle, $\om_0 = 0$.
Introducing $ X = x+y $, $ \xi = x-y$, it may be shown that
[\cite{CaL,HPZ}],
$$
J(X_f, \xi_f, t  | X_0, \xi_0 , 0 ) = {N \over \pi \hbar}
\ \exp\left(  \ih \tilde S - { \phi \over \hbar }  \right)
\eqno(2.19)
$$
where
$$
\tilde S= \tilde K(t) X_f \xi_f + \hat K(t) X_0 \xi_0
- L( t) X_0 \xi_f - N(t) X_f \xi_0
\eqno(2.20)
$$
and
$$
\phi  = A(t) \xi_f^2 + B(t) \xi_f \xi_0 + C(t) \xi_0^2
\eqno(2.21)
$$
Explicit expressions for the coefficients $A,B,C$ and
$ \tilde K, \hat K, L, N$ are given in the appendix.

\def\p{{\partial}}

\head{\bf 3. The Wigner Function Propagator}

Instead of the density operator, it is often convenient to work
with the Wigner function, defined by
$$
W(p,q) = { 1 \over 2 \pi \hbar} \int d \xi \ e^{-\ih p \xi}
\ \rho( q + \half \xi, q - \half \xi)
\eqno(3.1)
$$
Its inverse is,
$$
\rho(x,y) = \int dp \ e^{\ih p (x-y) } \ W ( p, {x+y \over 2} )
\eqno(3.2)
$$
(See Refs.[\cite{BaJ,Tat}] for properties of the Wigner function.)

The evolution of the
reduced density operator is described by the (generally non-unitary)
propagator, $J$, Eq.(2.4).
Correspondingly, one may introduced the Wigner function propagator,
$ K(p,q,t |p_0,q_0,0)$, defined by
$$
K(p,q,t|p_0,q_0,0) = {1 \over 2 \pi \hbar} \int d \xi d \xi_0
\ e^{-\ih p \xi}
J( q + \half \xi, q - \half \xi, t | q_0 + \half \xi_0, q_0 - \half
\xi_0, 0 )
\ \ e^{\ih p_0 \xi_0}
\eqno(3.3)
$$
and one has,
$$
W(p,q,t) = \int dp_0 dq_0 \ K(p,q,t | p_0, q_0, 0)
\ W(p,q,0)
\eqno(3.4)
$$

By performing the Wigner transform of the evolution equation for
$\rho$, Eq.(2.17), one may derive an analogous equation for the Wigner
function. It is
$$
\eqalignno{
{ \p W \over \p t} = &- { p \over M} { \p W \over \p q} +
V_R^{\prime}(q)
{ \p W  \over  \p p } + 2 \Gamma(t) { \p \over \p p} (p W) + \hbar
\Gamma(t) h(t) { \p^2 W \over \p p^2}
+ \hbar \Gamma(t) f(t) { \partial^2 W \over \partial q \partial p}
\cr
& + \sum_{k=1}^{\infty} \left( {i \hbar \over 2 } \right)^{2k} { 1
\over (2k+1)!} V^{(2k+1)}(q) \ { \p^{2k+1} W \over \p p^{2k+1} }
&(3.5) \cr }
$$
It follows that the Wigner function propagator also obeys this
equation, together with the initial conditions,
$$
K(p,q,0|p_0,q_0,0) = \delta (p-p_0) \ \delta (q-q_0)
\eqno(3.6)
$$

Consider the right-hand side of the Wigner function evolution
equation (3.5). The first two
terms are a Liouville operator, and if the equation had only these
terms, $W$ would evolve along the classical flow in phase space.
The third term is the dissipative term. It modifies the flow along
which $W$ evolves, and also causes a contraction of each area element.
The fourth and fifth terms are diffusive terms, and produce an
expansion of each area element. They are responsible
for noise/fluctuations, and also for the destruction of
interference, by erasing the structure of the Wigner function on
small scales. The last term, the power series, together
with the first two terms make up the unitary part of the evolution.
Hence, up to corrections of order $\hbar^2$, unitary evolution
corresponds to approximately classical evolution of the Wigner
function. It is partly for this reason that
the evolution of a quantum system is most conveniently undertaken in
the Wigner representation.
The higher corrections can often be argued to be
negligible, {\it e.g.}, if the Wigner function does not develop too
much detailed structure on small scales.
There are, however, important examples where they cannot be
neglected, {\it e.g.}, in chaotic systems [\cite{ZuP}].

To illustrate the simplicity of evolution in the Wigner
representation, we write down the solution for the propagator, in
the case in which the diffusive terms, dissipative terms
and higher order terms in $\hbar$ are ignored. It is,
$$
K(p,q,t|p_0,q_0,0) =
\ \delta (p - p^{cl}) \ \delta (q - q^{cl} )
\eqno(3.7)
$$
where $p^{cl} = p^{cl}(p_0,q_0,t)$ and $q^{cl} = q^{cl}(p_0,q_0,t)$
are the solutions to the classical equations of motion,
$$
\eqalignno{
\dot p &= -V'(q)
&(3.8)
\cr
\dot q &= { p \over M}
&(3.9)
\cr }
$$
satisfying the initial conditions $p^{cl} = p_0$, $q^{cl} = q_0$ at
$t=0$.

For the case of linear systems coupled to an environment, described
in the previous section, the density operator propagator is
given explicitly by (2.19)--(2.21). The Wigner
function propagator may be computed, yielding,
$$
\eqalignno{
K(p,q,t|p_0,q_0,0) & = { N \over \pi \hbar (4 AC -B^2)^{\half} }
\cr & \times
\ \exp \left( - \a (p - p^{cl} )^2 -\b (q- q^{cl})^2 -
\epsilon (p-p^{cl})(q-q^{cl}) \right)
&(3.10) \cr }
$$
where
$$
\eqalignno{
\a & = { C \over \hbar (4 AC - B^2) }
&(3.11) \cr
\b & = { 4 ( A N^2 + B N \tilde K + C \tilde K^2 ) \over
\hbar (4 AC - B^2) }
&(3.12) \cr
\epsilon & = - { 2 ( NB + 2 C \tilde K ) \over
\hbar ( 4 AC - B^2) }
&(3.13) \cr }
$$
and $q^{cl}$, $p^{cl} = M \dot q^{cl}$,
are the solution to the classical field
equations with dissipation and with a renormalized frequency,
$$
\ddot q + 2 \gamma \dot q + \omega^2_R q = 0
\eqno(3.14)
$$
matching $p_0$, $q_0$ at $t=0$. Explicitly,
$$
\eqalignno{
q^{cl} & = {p_0 \over 2N} + { \hat K \over N} q_0
&(3.15) \cr
p^{cl} & = { \tilde K \over N} p_0 - 2 (LN - \tilde K \hat K ) { q_0
\over N }
&(3.16) \cr }
$$

Now for what follows, it is important that the transformation from
$p_0,q_0$ to $p^{cl},q^{cl}$, defined by Eqs.(3.15), (3,16) is not a
canonical transformation, since the Jacobian of the transformation
is
$$
{ \partial (p^{cl},q^{cl}) \over \partial (p_0,q_0) } = { L \over N}
\eqno(3.17)
$$
This is because the classical evolution is dissipative, which tends
to cause phase space cells to shrink. This shrinking effect can be
compensated for by a scaling transformation of $p^{cl},q^{cl}$. In
particular, the variables,
$$
p' = \lambda^{-1} p^{cl}, \quad
q' = \lambda^{-1} q^{cl}
\eqno(3.18)
$$
where $\lambda = (L/N)^{\half}$,
{\it are} related to $p_0,q_0$ by a canonical transformation.
Using these new variables, the propagation of the Wigner function
may be written,
$$
\eqalignno{
W(p,q,t) & = \int dp_0 dq_0
\ { 1 \over \hbar (4 AC -B^2)^{\half} }
\cr  \times &
\exp \left( - \a (p - \lambda p')^2 -\b (q- \lambda q')^2 -
\epsilon (p-\lambda p' )(q- \lambda q') \right)
W(p_0,q_0,0)
&(3.19) \cr}
$$
Now one may perform the canonical transformation of integration
variables from $p_0,q_0$ to $p'q'$,
$$
\eqalignno{
W(p,q,t) & = \int dp' dq'
\ { 1 \over \hbar (4 AC -B^2)^{\half} }
\cr \times &
\exp \left( - \a (p - \lambda p')^2 -\b (q- \lambda q')^2 -
\epsilon (p-\lambda p' )(q- \lambda q') \right)
W'(p',q',0)
&(3.20) \cr}
$$
where
$$
W'(p',q',0) = W(p_0,q_0,0)
\eqno(3.21)
$$
Now the point is that the
quantity $W'$ defined by this transformation is still a Wigner function,
{\it i.e.}, it is the Wigner transform of density operator, $\rho'$,
say. In fact, it is readily shown that $\rho'$ is related to the
original $\rho$ by a unitary transformation. This would not be
true of the transformation from $p_0,q_0$ to $p^{cl},q^{cl}$.

Eq.(3.20) is the main result of this section: a simple expression
for the evolution of the Wigner function for linear systems from an
arbitrary initial state. The (non-dissipative part of the) classical
evolution has been absorbed into a canonical transformation of the
initial state, and the effects of dissipation and diffusion are
contained in the coefficients, $\a,\b,\epsilon,\lambda$.

It is sometimes convenient to write the Wigner function propagator
in the alternative form,
$$
\eqalignno{
K(p,q,t|p_0,q_0,0) & =  { N  \over \pi \hbar (4 AC -B^2)^{\half} }
\cr  \times &
\ \exp \left( - \mu (p_0 - p_0^{cl} )^2 -\nu (q_0- q_0^{cl})^2 -
\sigma (p_0-p_0^{cl})(q_0-q_0^{cl}) \right)
&(3.22) \cr }
$$
where
$$
\eqalignno{
\mu & = { A \over \hbar (4 AC - B^2) }
&(3.23) \cr
\nu & = { 4 ( A \hat K^2 + B L \hat K + C L^2 ) \over
\hbar (4 AC - B^2) }
&(3.24) \cr
\sigma & =  { 2 ( LB + 2 A \hat K ) \over
\hbar ( 4 AC - B^2) }
&(3.25) \cr }
$$
$p_0^{cl}$, $q_0^{cl}$ are again solutions to the classical field
equations with dissipation,
but now matching the final conditions $p$, $q$ at time $t$:
$$
\eqalignno{
q_0^{cl} & = - {p \over 2L} + { \tilde K \over L} q
&(3.26) \cr
p_0^{cl} & = { \hat K \over L} p - 2 (LN - \tilde K \hat K ) { q
\over L }
&(3.27) \cr }
$$
We will use these expressions in the following sections.

\head{\bf 4. Generalized Uncertainty Relations}

We now show how the results of the previous section may be used to
derive generalized uncertainty relations for quantum Brownian
motion models, for linear systems.

\subhead{\bf 4(A). A Lower Bound on the Uncertainty Function}

{}From the Wigner function propagator, one may obtain expressions for
the distributions of $p$ and $q$ at time $t$:
$$
\eqalignno{
\rho(q,q,t) &= \int dp \ W(p,q,t)
\cr
&= \int dp_0 dq_0
{ 1 \over (2 \pi s_q^2)^{\half} }
\ \exp
\left( - { (q- \lambda q')^2 \over 2 s_q^2}
\right) \ W'(p',q',0)
&(4.1) \cr
\tilde \rho(p,p,t) & = \int dq \ W(p,q,t)
\cr
& =\int dp_0 dq_0
{ 1 \over (2 \pi s_p^2)^{\half} }
\ \exp \left(
- { (p- \lambda p')^2 \over 2 s_p^2}
\right) \ W'(p',q',0)
&(4.2) \cr }
$$
where
$$
s_q^2 = {2 \a \over ( 4 \a \b - \epsilon^2) }, \quad
s_p^2 = {2 \b \over ( 4 \a \b - \epsilon^2) }
\eqno(4.3)
$$
{}From these results, it is straightforward to compute the variances
of $p$ and $q$ at any time $t$:
$$
\eqalignno{
(\Delta q_t)^2 &= \lambda^2 ( \Delta q' )^2 + s_q^2
&(4.4) \cr
(\Delta p_t)^2 &= \lambda^2 ( \Delta p' )^2 + s_p^2
&(4.5) \cr }
$$
where $ ( \Delta q' )^2 $, $( \Delta p' )^2 $ denote the
variances of $q$ and $p$ in the canonically transformed initial
Wigner function (3.21), or equivalently, in the unitarily transformed
initial density operator $\rho_0'$. (The variances are generally not
invariant under such transformations of the state.)

We have the usual uncertainty principle,
$$
(\Delta q') (\Delta p') \ge { \hbar \over 2}
\eqno(4.6)
$$
with equality if and only if $\rho'$ is a coherent state, {\it
i.e.}, if and only if
$$
W'(p',q',0) = { 1 \over \pi \hbar} \ \exp \left(
- {2 (\D q')^2 \over \hbar^2 } {p'}^2 - { {q'}^2 \over 2 (\D q')^2 } \right)
\eqno(4.7)
$$
(where $\Delta q'$ is, so far, an arbitrary parameter).
It follows that the uncertainty function at time $t$ satisfies,
$$
U_t \ \equiv \ (\Delta q_t)^2 (\Delta p_t)^2 \ \ge
\ { \hbar^2 \lambda^2 \over 4} +  \lambda (\Delta q')^2 s_p^2
+ { \hbar^2 \lambda s_q^2 \over 4 (\Delta q')^2 } + s_q^2 s_p^2
\eqno(4.8)
$$
with equality if and only if the initial state is given by (4.7).
Since $\Delta q'$ is arbitrary, we may minimize over it. The minimum
is at
$$
(\Delta q')^2  = { \hbar s_q \over 2 s_p }
\eqno(4.9)
$$
and inserting the minimum value in (4.7), we obtain
$$
(\Delta q_t) (\Delta p_t) \ \ge
\ { \lambda \hbar \over 2} + s_p s_q
\eqno(4.10)
$$
In terms of the coefficents of the non-unitary propagator,
(3.11)--(3.13), this reads,
$$
(\Delta q_t) (\Delta p_t) \ \ge
{ \hbar L \over 2 N} + { \hbar \over 2 N^2}
\left( AC N^2 + BC N \tilde K + C^2 \tilde K^2 \right)^{\half}
\eqno(4.11)
$$
This the first main result of this section: the uncertainty at time $t$
satisfies the generalized uncertainty relation (4.11), with equality
if and only if the initial state is given by (4.7) with $\Delta
q'$ given by (4.9).

Consider now the conditions for equality. From (4.7) and (3.21), the
minimizing initial Wigner function is
$$
W(p_0,q_0,0) = { 1 \over \pi \hbar} \ \exp\left( - {2 \s^2 \over
\hbar^2 LN} ( \tilde K p_0 - 2 (LN - \tilde K \hat K) q_0 )^2
- { 1 \over 2 \s^2 LN} (\half p_0 + \hat K q_0 )^2 \right)
\eqno(4.12)
$$
Inverting the  Wigner transform, one finds that the minimizing state
is a pure state,
$$
\Psi(x) = \left( { 2a \over \pi } \right)^{1/4}
\ \exp \left( - (a+ib) x^2 \right)
\eqno(4.13)
$$
This is a so-called correlated coherent state [\cite{DKM}],
discussed below. The coefficients $a$ and $b$ may be computed from
(4.12) but will not be needed here so are not given.

It is important to note that the initial state (4.13) minimizing the
uncertainty at time $t$ is {\it different for each moment of time}.
That is, the lower bound (4.11), is {\it not} the time evolution of a
particular initial state. It is actually an envelope. The initial
state (4.13), which depends on $t$, will minimize the uncertainty at
time $t$ but generally not at other times. This is because, although
the effect of diffusion is generally to increase the uncertainty as
times goes on, there are competing effects that may reduce it.
In particular, wave packet reassembly (the time reverse of wave
packet spreading) and dissipation may cause the uncertainty to
decrease.
This point is discussed further in Ref.[\cite{AnH}], where
an analagous lower bound on the Wehrl entropy (discussed below) was
proved.

\noindent {\bf 4(B). A Sharper Lower Bound}

As stated,
the lower bound on the uncertainty (4.11) is sharp, in the sense that
at each moment of time there exists an initial state at which the
lower bound is achieved. But it is only an envelope because the
the initial state achieving it is different for each moment of time.
It turns out that an improvement of this result is possible.

First, we note that the the conventional uncertainty relation,
$ \Delta p \Delta q \ge {\hbar /2 } $, may be generalized to the
stronger result [\cite{DKM}],
$$
{\cal A}^2 \equiv
(\Delta p)^2 (\Delta q)^2 - C_{pq}^2 \ge { \hbar^2 \over 4}
\eqno(4.14)
$$
where $C_{pq}$ is the correlation between $p$ and $q$,
$$
\eqalignno{
C_{pq} =& \half\  \langle \D \hat p \D \hat q + \D \hat q
\D \hat p \rangle
\cr
=&  \int dp dq \ p q \ W(p,q) - \la \hat p \ra \la \hat q \ra
&(4.15)
\cr }
$$
where $\D \hat q = \hat q - \la \hat q \ra $, $\D \hat p -
\hat p = \la \hat p \ra $. The quantity ${\cal A}$ defined by (4.14)
is essentially the area enclosed by the $1-\sigma$ contour of the Wigner
function.
Equality in (4.14)
is achieved when the state is a correlated coherent state,
$$
\Psi(x) = { 1 \over (2 \pi \eta^2)^{1/4} } \ \exp \left( - {x^2
\over 4 \eta^2} \left[ 1 - {i r \over (1-r^2)^{\half}} \right] +
{ \a x \over \eta} - \half ( \a^2 + | \a |^2 ) \right)
\eqno(4.16)
$$
where $ \eta, r $ are real parameters ($ r = C_{pq} / (\Delta p
\Delta q) $ and $ \eta = \Delta q $), and $\a$ is a complex parameter.

We have, however, already seen a correlated coherent state -- it is
the state minimizing the uncertainty relation (4.11). It is therefore
plausible that a stronger version of (4.11) might be possible in
terms of the Wigner function area (4.14). Indeed,
computing the correlation from (3.19), one finds the remarkably
simple result,
$$
{\cal A}_t^2 \ = \
{ L^2 \over N^2 } {\cal A}_0^2
+ { \hbar^2 \over 4 N^2 } ( 4 AC - B^2 )
\eqno(4.17)
$$
This gives the Wigner function area ${\cal A}$
at an arbitrary time $t$, ${\cal A}_t$,
in terms of its value at time
$t=0$, ${\cal A}_0$.
It is remarkable because ${\cal A}$ at time $t$ depends
on {\it only} its value at $t=0$, and not on any of the other
moments of the initial state.
Now at $t=0$, ${\cal A}_0$ satisfies the
uncertainty principle (4.14). We thus obtain, for the Wigner
function area at time
$t$, the inequality
$$
{\cal A}_t^2 \ \ge \
{ \hbar^2 L^2 \over 4 N^2 }
+ { \hbar^2 \over 4 N^2 } ( 4 AC - B^2 )
\eqno(4.18)
$$
Equality is achieved when the initial state belongs to the four-parameter
family of correlated coherent states. Most importantly, the
inequality is sharp at every moment of time for one and the same
initial state. It is not an envelope. Eq.(4.18) is the second main result
of this section.

The simplicity of the time evolution in terms of the the Wigner
area is easy to
understand.
For linear systems,
the area is preserved under unitary time evolution. For the
non-unitary case considered here, therefore, (4.17) expressess
entirely the effects of the environment, with the effects of unitary
evolution completely factored out. In this respect it is similar to
the von Neumann entropy discussed in the next section. Indeed, for Gaussian
states, the von Neumann entropy is a function only of the Wigner area.

Some rather different, but not unrelated generalizations of the
uncertainty principle may be found in Refs.[\cite{Hal2,Hal3}].

\head {\bf 5. Explicit Forms of the Lower Bounds}

We now give the explicit forms of the lower bounds (4.11) and (4.18), in a
variety of situtations. There are a very large number of different situations
and regimes
that our results may cover, depending on the choice of $t$, $T$, $\gamma $
and $\omega$ (harmonic oscillator in the under- and over-damped case,
inverted harmonic oscillator, free particle). We will not present here a
completely exhaustive search of the parameter space, but concentrate on
the cases that contain some interesting physical results. In
particular, most of the following results are the underdamped case.
The cases
not covered here are readily derived, should the reader be interested.

Consider first the case of short times, $ t << \gamma^{-1} $. Then
for high temperatures, (4.11) and (4.18) become, respectively,
$$
(\D q_t ) ( \D p_t )
\ \ge \ { \hbar  \over 2} \left( 1 - 2 \gamma t + { 8 \sqrt{3} \over 3}
{ \gamma k T \over \hbar } t^2 \right)
\eqno(5.1)
$$
$$
\A^2_t \ \ge \ { \hbar^2 \over 4}
\left( 1 - 4 \gamma t + { 8 \over 3 } { \gamma^2 k^2 T^2 \over
\hbar^2 } t^4 \right)
\eqno(5.2)
$$
These relations are in fact valid for any potential.
They represent the initial growth of
fluctuations, starting from the purely quantum fluctuations at $t=0$.

The  $\gamma t $ term in each expression indicates an initial decrease
of fluctuations, in apparent violation of the uncertainty principle.
This violation occurs, in Eq.(5.1) for example,
on a timescale less than $ \hbar / kT $.
This is because these expressions have been derived by taking the
infinite cut-off limit in Eq.(2.13), which as previously stated,
can lead to violations of positivity of the density operator on
time scales less than $\Lambda^{-1}$ [\cite{Amb}].
The expressions are therefore not
valid for $t< \Lambda^{-1}$. The high temperature limit
used in deriving (5.1) means $ k T >> \hbar \Lambda $. Combining these
inequalities, we see that (5.1) is valid only for times
$t> \Lambda^{-1} >> { \hbar / k T} $, for which violations of
the uncertainty principle will not arise.

To illustrate explicitly that it is the infinite cutoff limit that
is responsible for violations of the uncertainty principle, we
compute the lower bound (4.11) leaving the cutoff finite in the
expressions for the various coefficients in the propagator
(2.19)--(2.21). The important point is to keep $\Lambda$ finite in
Eq.(2.12). Expressions (A.1)--(A.4) and (A.14) are no longer valid,
and the correct expressions for these coefficients, in the short
time limit, are given by Eqs. (A.28)--(A.32). One thus obtains, in the
short time limit, the lower bound
$$
( \D q_t ) ( \D p_t ) \ \ge \ { \hbar \over 2} + { \gamma k T
\Lambda \over 2 \pi } t^3 + ...
\eqno(5.3)
$$
for which there is clearly no violation of the uncertainty principle.

Ignoring the positivity-violating terms in (5.1), (5.2),
the remaining terms give an indication of the time scale on which the
thermal fluctuations become important in comparison to the quantum ones.
It is,
$$
t \sim \left( {\hbar \over \gamma k T } \right)^{\half}
\eqno(5.4)
$$
As noted in Ref.[\cite{BrK}], this is the timescale on which quantum
fluctuations become comparable to Nyquist noise.

The timescale (5.5) differs from that derived in
Refs.[\cite{HuZ,AnH}], where it was argued that thermal fluctuations
become comparable to quantum ones on the timescale
$ t_d \sim \hbar^2 / (M \gamma k T \ell^2) $ (the timescale
characteristic of decoherence [\cite{Zur}]). Here $\ell$ is a
length scale, arising from the particular way in which the
fluctuations were measured in Refs.[\cite{HuZ,AnH}] ({\it e.g.}, in
Ref.[\cite{AnH}] is was the width of a phase space projector).
There is no
length scale present in the method use to measure fluctuations
here, hence a different result is obtained. The exact connection
between
these two results is not very clear. We note, however, that the time
scale (5.4) will typically be much longer than the decoherence time
scale.

In the weak coupling regime, for all temperatures and times (except very short
times), Eq.(4.18) becomes, using (A.18)--(A.20),
$$
A_t^2 \ge { \hbar^2 \over 4 } \left[ e^{ -4 \gamma t} + \coth^2 \left(
{ \hbar \om \over 2 k T } \right) ( 1 - e^{-2 \gamma t} )^2 \right]
\eqno(5.5)
$$
An expression may also be derived for arbitrary coupling and high temperatures,
using (A.11)-(A.13), but this is rather complicated and will not be given.

In the long-time limit, for any coupling and temperature,
$$
U_t = \A^2_t \ge { \hbar^2 \over \om_R^2}
\left(F_1^2 - { \gamma^2 \over \om^2} F_2^2 \right)^{\half}
\eqno(5.6)
$$
In the high temperature limit this becomes,
$$
U_t = \A^2_t \ge  { k^2 T^2 \over \om_R^2 }
\eqno(5.7)
$$
and in the weak coupling limit,
$$
U_t = \A^2_t \ge { \hbar^2 \over 4} \coth^2 \left( {\hbar \om \over 2 k T}
\right)
\eqno(5.8)
$$

\head {\bf 6. Inequalites Relating Uncertainty to Entropy and
Linear Entropy}

The von Neumann entropy of the reduced density operator,
$$
S[\rho] = - {\rm Tr} ( \rho \ln \rho)
\eqno(6.1)
$$
is often discussed in the present context, in association with
uncertainty, decoherence, and correlations of the distinguished
systems with its environment [\cite{PHZ,PaZ,ZHP,Zur2,JoZ}].
Zurek, Paz, and Habib
for example, looked for classes of initial
states, which under evolution according to a master equation of the
form (2.17), generate the least amount of entropy at time $t$. They
regarded such states as the most stable under evolution in
the presence of an environment. They argued that the initial states
doing the job are coherent states, at least approximately.

One of the reasons for looking at the von Neumann entropy is that it is
constant for unitary evolution, thus for open systems such as those
considered here, it is principally a measure of environmentally
induced effects. The Wigner area
considered in Section IV also has this property.
The results of Section IV thus agree
with the claims of Zurek {\it et al.} (except that it is
really the correlated coherent states, rather than the ordinary
coherent states which are the most stable, although this distinction
is not very important).
It would be useful to find a connection between these two measures
of stability or fluctuations.

\subhead{\bf 6(A). Uncertainty vs. von Neumann Entropy}

The connection between uncertainty and the von Neumann entropy
may be found indirectly, by considering first of all
the phase space distribution,
$$
\mu (p,q) = \la z | \rho | z \ra
\eqno(6.2)
$$
where
$$
\la x |z \ra = \la x |p,q\ra = \left( {1\over 2 \pi \s_q^2}\right)^{1/4}
\exp \left(-{(x-q)^2\over 4\s_q^2} +\ih p x  \right)
\eqno(6.3)
$$
are the standard coherent states.
The function $\mu(p,q)$ is normalized according to
$$
\int { dp dq \over 2 \pi \hbar} \ \mu(q,p)=1.
\eqno(6.4)
$$
It is readily shown that $\mu(p,q)$ is also equal to
$$
\mu (p,q) = 2 \int dp' dq' \ \exp
\left( - { (p-p')^2 \over 2 \s_p^2 } - { (q-q')^2
\over 2 \s_q^2 } \right) \ W_{\rho} (p',q')
\eqno(6.5)
$$
where $ W_{\rho} (p,q) $ is the Wigner function of $\rho$
and $ \s_p \s_q = \half \hbar $.
Eq.(6.5) is sometimes known as the Husimi distribution [\cite{Hus}],
and is positive even though the Wigner function is not in general
(see also Ref.[\cite{Hal1}]).

There exists an information-theoretic
measure of the uncertainty or spread in phase space contained in
the distribution (6.5), namely the so-called Wehrl entropy [\cite{Weh}]
$$
I(P,Q) = - \int { dp dq \over 2 \pi \hbar} \ \mu(p,q) \ln \mu(p,q)
\eqno(6.6)
$$
This is the Shannon information of (6.2). $I(P,Q)$ is large for
spread out distributions, and small for very concentrated ones.
Because of the uncertainty principle, one would expect a limit on
the degree to which $\mu(p,q)$ may be concentrated about a small
region of phase space, and hence a lower bound on (6.6).
Indeed, a non-trivial theorem due to Lieb shows that,
$$
I(P,Q)  \ \ge \ 1
\eqno(6.7)
$$
with equality if and only if $ \rho $ is the density matrix of
a coherent state, $ |z' \ra \la z' | $ (Ref.[\cite{Lie}]).
In Ref.[\cite{AnH}], the Wehrl
entropy was used as a measure of both quantum and thermal
fluctuations, and a lower bound analagous to (4.11) was derived.

The reason for studying this quantity is that it provides the
link between the von Neumann entropy and
the uncertainty measures $U$ and $\A$. On the one hand,
an elementary property of Shannon information is
$$
I(P,Q) \ \le \ \ln \left( { e \over \hbar} ({\rm det} K)^{\half}
\right)
\eqno(6.8)
$$
where $K$ is the $2 \times 2$ covariance matrix of the distribution
$\mu(p,q)$ [\cite{Cov}].
Equality holds if and only if $\mu(p,q)$ is a Gaussian.
{}From (6.5), one has
$$
{\rm det} K =
\left( ( \D  q )^2 + \s_q^2 \right) \left( (\D  p )^2 + \s_p^2 \right)
- C_{pq}^2
\eqno(6.9)
$$
where $\D q$, $\D p$ are the quantum-mechanical variances
of the state $\rho$.
On the other hand, an elementary property of the Wehrl entropy,
following from the concavity of Shannon information is [\cite{Weh}],
$$
I (P,Q) \ \ge \ S[ \rho]
\eqno(6.10)
$$
Hence combining the upper and lower limits (6.8), (6.10), one
obtains,
$$
\left( (\D q )^2 + \s_q^2 \right) \left(( \D p )^2 + \s_p^2 \right)
- C_{pq}^2 \ \ge \ \hbar^2 e^{2(S-1)}
\eqno(6.11)
$$
Finally, since the parameter $\s_q$ is arbitrary (and $\s_p \s_q =
\hbar/2 $), we may minimize the left-hand side over it, with the
result,
$$
( \D p \D q + \half \hbar )^2 - C_{pq}^2 \ \ge \ \hbar^2 e^{2(S-1)}
\eqno(6.12)
$$
This is the exact form of the connection between the uncertainty and
entropy for a general mixed state $\rho$.

In the regime where quantum fluctations are more significant than
thermal ones, it is appropriate to use the lower bound (6.7) rather
than (6.10) (since $S[\rho]$ goes to zero if the
state is pure), and this is formally achieved by setting $S=1$ in
(6.12). One then deduces the usual uncertainty principle from (6.12)
(although not the generalized version including the correlation,
(4.14)).

In the regime where thermal (or environmentally-induced)
fluctuations are dominant, one would expect
$\D p \D q >> \hbar/ 2 $ and $ S >>1 $, and (6.12) then gives
$$
 { \D p  \D q \over \hbar } \ \ge
\ { {\cal A} \over \hbar } \ \ge \ e^{S}
\eqno(6.13)
$$
This is the simplest form of the connection between the
uncertainty and the von Neumann entropy: the entropy is
bounded from above by the logarithm of the number of phase space
cells the state occupies.

To see how sharp these equalities can be, consider the
case of a Gaussian Wigner function. It may be shown
that the von Neumann entropy of a Gaussian is
$$
S[\rho] = - \ln \left( {2 \over 1 + \mu} \right)
- { ( \mu - 1) \over 2 } \ln \left( { \mu - 1 \over
\mu + 1 } \right)
\eqno(6.14)
$$
where $\mu = 2 {\cal A} / \hbar $ and ${\cal A}$ is the
area of the Gaussian Wigner function (see Ref.[\cite{JoZ}]
for example). For large ${\cal A}$,
$$
S [\rho] \approx \ln \left( { { \cal A} \over \hbar } \right)
\eqno(6.15)
$$
and hence we have equality in (6.13).

\subhead{\bf 6(B). von Neumann Entropy vs. Linear Entropy}

In practice, the discussions of Zurek {\it et al.}, and indeed many
discussions of von Neumann entropy, often concern the so-called
linear entropy,
$$
S_L = 1 - {\rm Tr} \rho^2
\eqno(6.16)
$$
How is this related to von Neumann entropy? Let the density operator
be
$$
\rho = \sum_n \ p_n \ |n \rangle \langle n |
\eqno(6.17)
$$
We use Jensen's inequality (see, Ref.[\cite{Cov}], for example),
which states
that if $f$ is a convex function, and $X$ a random variable,
$$
\langle f(X) \rangle \ \ge \ f ( \langle X \rangle )
\eqno(6.18)
$$
where $ \langle \cdots \rangle $ denotes the mean over $X$.
Take $f$ to be the exponential function, and $X$ to be $\ln p_n$.
Then,
$$
\sum_n p_n^2 \ \ge \ \exp \left( \sum_n p_n \ln p_n \right)
\eqno(6.19)
$$
and thus
$$
{\rm Tr} \rho^2 \ \ge \ e^{-S[\rho]}
\eqno(6.20)
$$
In terms of the linear entropy,
$$
S_L = 1 - {\rm Tr} \rho^2 \ \ge \ 1 - e^{-S}
\eqno(6.21)
$$
Equality in (6.21) is reached for pure states, when $S = S_L = 0 $, and
is for very mixed states, when $S$ is very large and $S_L \approx  1$.

\head {\bf 7. Long-Time Limits}

One of the particularly interesting questions for non-equilibrium
systems of the type considered here is whether they settle down to
a unique state after a long period of time. It turns out to
be particularly straightforward to answer this question for linear
systems, using the Wigner function propagator described in Section III.
To see this, note that combining (3.4) and (3.10), we have the
expression,
$$
\eqalignno{
W(p,q,t) = & { N  \over \pi \hbar (4 AC -B^2)^{\half} }
\int dp_0 dq_0 \ W(p_0, q_0, 0 )
\cr & \times
\ \exp \left( - \a (p - p^{cl} )^2 -\b (q- q^{cl})^2 -
\epsilon (p-p^{cl})(q-q^{cl}) \right)
&(7.1) \cr }
$$
where, recall, $p^{cl}, q^{cl} $ are the value of the
classical solutions (with dissipation)
at time $t$ matching the initial data
$ p_0 $, $q_0 $ at $t=0$. This expression allows us to compute
the moments at any time $t$ in terms of the initial moments.
By computing the long-time limits of these moments, the form of the
long-time limit of the Wigner function may be obtained, since it is
completely determined by its moments.

For the harmonic oscillator, simplifications occur.
Consider first the case of the harmonic oscillator in an
ohmic envrionment.
One has $p^{cl} = M \dot q^{cl}$ and $q^{cl}$ satisfies
Eq.(3.14). Either from Eq.(3.14),
or from the explicit solution (3.15), (3.16), it is easily seen that
$ p^{cl} \ria 0$ and $q^{cl} \ria 0 $ in the long time limit.
All dependence on $p_0$ and $q_0$ drops out of the exponential
in (7.1), and one obtains the following expression for the
the asymptotic value of the Wigner function,
$$
W_{\infty} (p,q) ={ N \over \pi \hbar (4 AC -B^2)^{\half} }
\exp \left( - \a p^2 -\b q^2 -
\epsilon p q  \right)
\eqno(7.2)
$$
The coefficients $\a, \b, \epsilon $ are given by the long-time limits
of (3.11)--(3.13). Using results (A.21)--(A.23) in the appendix, one finds
$$
\eqalignno{
\a = & { 1 \over 2 M \hbar (F_1 - { \gamma \over \om } F_2) }
&(7.3) \cr
\b = & { M \om_R^2 \over 2 \hbar (F_1 + { \gamma \over \om } F_2 ) }
\cr
\epsilon = & 0
&(7.4) \cr }
$$
with $F_1$ and $F_2$ given by (A.24), (A.25). These relations represent
the {\it exact} form of the long-time limit in the ohmic case.

It is useful to compare the result (7.2) with the Wigner function of
the harmonic oscillator in a thermal state:
$$
W_T (p,q) = { 1 \over \pi \hbar} \tanh \left( {\hbar \om_R \over 2 kT}\right)
\exp \left( - \tanh \left( {\hbar \om_R \over 2 kT}\right)
\left[ { M \om_R \over 2 kT} q^2 + { 1 \over M \hbar \om_R} p^2 \right]
\right)
\eqno(7.5)
$$
The long-time limit (7.2) coincides with the thermal Wigner function (7.5)
in the high-temperature limit, and in the weak coupling limit, as may be
seen from Eqs.(A.26), (A.27) in the appendix. Note, however, that the
asymptotic
state is not always a thermal state.

It is also of interest to compute the uncertainty in $q$ in the long
time limit. It is,
$$
\eqalignno{
( \D q )^2_{\infty} & = { 2 \hbar \over M \omega_R^2 } \left( F_1 + { \gamma
\over \omega } F_2 \right)
\cr
& = { \hbar \over \pi}
\int_0^{\infty} d \nu \ \coth \left( { \hbar \nu \over 2
k T} \right) \ { 1 \over M} \ { 2 \gamma \nu \over ( \omega_R^2 -
\nu^2 )^2 + 4 \gamma^2 \nu^2 }
& (7.6) \cr }
$$
This is in agreement with the fluctuation-dissipation relation [\cite{Lan}].
Eq.(7.6) was given in Ref.[\cite{CaL}], for the special case of a
Gaussian initial state, but not surprisingly coincides
with this more general result because every initial state goes to a Gaussian.

For the harmonic oscillator with a non-ohmic environment, it may be shown,
using the more general treatment of Ref.[\cite{HPZ}], that $q^{cl}$ in the
Wigner function propagator in (7.1) is the solution to the
integro-differential equation,
$$
\ddot q (t) + \om_0^2 q (t) - \int_0^t ds \ \eta(t-s) \ q(s) = 0
\eqno(7.7)
$$
where $\eta(t-s)$ is dissipation kernel (2.10) for an arbitrary
spectral density $I(\omega)$. We conjecture that there exists
a large class of spectral densities for which
all solutions to (7.7)
will satisfy $q^{cl} \ria 0$, $ p^{cl} \ria 0 $ in the
long-time limit, although we have not been able to prove this.
If this conjecture is true, then the long-time limit of the Wigner function
will again be of the form (7.2) (although the coefficients $\a$, $\b$,
$\epsilon$ will not necessarily be the same).

These results, including the above conjecture, are consistent with the proof
in Ref.[\cite{TeS}], that the Wigner function of
a single member of a chain of coupled harmonic oscillators tends
towards a Gaussian Wigner function in the long-time limit, under
certain reasonable conditions on the environment.

\def\L{{\cal L}}
\head {\bf 8. General Potentials and the Free Particle}

All of the results we have described so far concern the specific case of
linear systems. It is clearly of interest to generalize our considerations
to other types of potential, and this we now do.

We are interested in the propagator (2.4) for arbitrary potentials
$V(x)$. It is generally not possible to evaluate this propagator for
arbitrary $V(x)$, so some approximations need to be made. Again it
is most convenient to work with the Wigner function evolution
equation (3.5). The coefficients of the dissipative and diffusive
terms in (3.5) are known only in either the linear case treated
earlier, or in the high-temperature limit, so we consider the
latter. We will also  neglect the infinite power series in
derivatives of $W$ and $V$, and hence the equation becomes,
$$
{\p W
\over \p t} = -{ p \over M} { \p W \over \p q } + V_R'(q) {\p W
\over \p p} + 2 \gamma { \p \over \p p} ( pW) + 2 M \gamma k T {
\p^2 W \over \p p^2}
\eqno(8.1)
$$
We expect that the neglect of
higher derivative terms in $W$ and  $V$ is valid when the initial
Wigner function does not possess, or develop,  too much detailed
structure in regions of size $\hbar$ or less. It is argued in
Ref.[\cite{ZuP}] that the  diffusive terms rapidly smooth the Wigner
function on small scales, and thus (8.1) is plausibly valid at long
times for any initial state.

As in the linear case, it is possible to associate a Wigner function
propagator (3.4), with (8.1), although it is not possible to calculate
it explicitly. We have explored the possibility that the Wigner function
propagator is given approximately by an expression of the form
(3.10), where $p^{cl}, q^{cl}$ are the solutions to the dissipative
equations of motion with potential $V(x)$, and the coefficients
$\a,\b,\epsilon$ are determined by substituting (3.10) into (8.1).
Unfortunately we have not been able to justify this approximation,
and as a consequence, we have not extended our results on the time-dependent
lower bounds on uncertainty functions derived in Section IV
to arbitrary potentials.
We will therefore work directly with Eq.(8.1), and consider only the
question of the long-time limit.

\subhead{\bf 8(A). Long Time Limits for General Potentials}

Eq.(8.1) is the Kramers equation. Its properties are described in some
detail in the book by Risken [\cite{Ris}], and we shall
make use of those results here. Many of the results in Risken
concern non-negative phase space
distribution functions, but we have been
careful not to assume that here, because the Wigner function is of
course not always positive.

It is easy to see that the master equation possesses the stationary solution,
$$
W(p,q) = \tilde N \exp \left( - { p^2 \over2 M k T} - { V_R(q) \over k T}
\right)
\eqno(8.2)
$$
where $\tilde N$ is a normalization factor.
This will be an admissable solution, {\it i.e.}, a Wigner function,
only if the potential is such that $\exp \left( - V_R(q) / kT \right) $
is normalizable. This requires $V_R(q) \ria \infty $ as $ q \ria \pm
\infty $ faster than $\ln |q|$. In that case, the stationary distribution is
then the Wigner transform of a thermal state, $ \rho = Z^{-1} e^{-H /kT}$
where $Z = {\rm Tr} ( e^{ - H/kT} ) $, for large
temperatures\footnote{$^{\dag}$}{Actually, the only Wigner function that
can be positive is a Gaussian
one [\cite{Tat}]. However, there is no contradiction with (8.2) because
we are working with a distribution function that satisfies
the full Wigner function equation only up to order $\hbar^2$,
so the distribution function is not exactly a Wigner function}.

The master equation may be written
$$
- {\p W \over \p t} = \L W
\eqno(8.3)
$$
where $\L$ is an operator deducible from (8.1), and acts on the set of
square integrable functions on phase space, with the inner product
between functions defined by
$$
(f,g) = \int dp dq \ f(p,q) \ g(p,q)
\eqno(8.4)
$$
The general solution to the master equation may then be written
$$
W(p,q,t) = \sum_n \ c_n \ e^{- \lambda_n t}\ \phi_n(p,q)
\eqno(8.5)
$$
where the $c_n$ are constants, and
$\phi_n$ are the eigenfunctions of $\L$, with eigenvalue
$\lambda_n$,
$$
\L \phi_n = \lambda_n \phi_n
\eqno(8.6)
$$
It may be shown that if the master equation admits a stationary solution,
{\it i.e.}, if one of the eigenvalues is zero, then the real part of the
remaining eigenvalues is strictly positive [\cite{Ris}]. It immediately
follows that
the general solution will tend towards the unique stationary
solution (8.2) in the limit $ t \ria \infty $. Hence every initial
state tends to the thermal state in the long time limit.

For potentials which are not bounded from below, the long-time limit
will often depend quite sensitively on the initial state. Indeed, the
inverted oscillator was recently used by Zurek and Paz as a prototype
model for chaotic systems [\cite{ZuP}].

\subhead{\bf 8(B). Long Time Limit for the Damped Free Particle}

Finally, let us consider the case of the free particle, {\it i.e.},
$V_R(q) = 0 $ in Eq.(8.1). In this case, there is no stationary
solution, because
the operator $\L$ does not have any zero eigenvalues. To see this,
let us look for eigenfunctions of the form
$$
u_{nk} (p,q) = e^{ikq} f_n (p)
\eqno(8.7)
$$
In Ref.[\cite{Ris}], it is shown that the eigenvalue equation for a
zero eigenvalue, $\L u_{nk} = 0$, is satisfied only if $k=0$. Hence
although a formal solution to the eigenvalue equation exists, it
does not belong to the spectrum of $\L$ because it is independent of
$q$ and so will not be normalizable.

Despite the absence of a stationary solution,
it is still of interest to ask whether the one can say anything at
all about the long time limit of the solution. Consider Eq.(7.1) for
the damped free particle. For large $t$, one  has
$$
\eqalignno{
\a &= {1 \over 2 M kT}, \quad
\b = { M \gamma \over 2 k T t}, \quad
\epsilon = - { 1 \over 2 k T t}
&(8.8) \cr
A &= B = { M k T \over 2 \hbar}, \quad C = { 2 M k T \gamma t \over
\hbar}
&(8.9) \cr
p^{cl} &= 0, \quad q^{cl} = q_0 + {p_0 \over 2 M \gamma}
&(8.10) \cr }
$$
Introducing $\tilde q = q_0 + p_0 / ( 2 M \gamma) $, Eq.(7.1) may be
written,
$$
\eqalignno{
W(p,q,t)  =  { N \over \pi \hbar (4 AC -B^2)^{\half} }
&  \exp \left( - \left( \a - { \e^2 \over 4 \b} \right) p^2 \right)
\cr \times
\int d \tilde q \ g_0 (\tilde q)
\ & \exp \left( - \b \left( q + { \e \over 2 \b} p - \tilde q \right)^2
\right)
&(8.11) \cr}
$$
where
$$
g_0(\tilde q) = \int dp_0 \ W (p_0, \tilde q  - { p_0 \over 2 M \gamma }, 0 )
\eqno(8.12)
$$
The integrand of (8.12) is still a Wigner function, since the shift in the
$q$ argument can be compensated for by a unitary transformation of the
density operator.

If we integrate out $q$, and noting that $ \e^2 / 4 \b \ria 0 $ for
large
$t$, we obtain
$$
\int dq \ W(p,q,t) = { 1 \over (2 \pi M k T)^{\half} }
\ \exp \left( - { p^2 \over 2 M k T } \right)
\eqno(8.13)
$$
Hence the distribution of momenta approaches a Boltzmann
distribution for all initial states.
The remaining question is, what we can say
about the $q$ distribution in the long time limit?

Let $ y = q + { \e \over 2 \b } p = q - { p \over 2 M \gamma }$.
Then the integral in the expression for the Wigner function (8.11) is,
$$
\eqalignno{
g(y,t)   \ &  \equiv \  \int d \tilde q \ g_0 (\tilde q)
\ \exp \left( - \b \left( q + { \e \over 2 \b} p - \tilde q \right)^2
\right)
\cr
& =
\int d \tilde q \ g_0 (\tilde q)
\ \exp \left( - { ( y - \tilde q)^2 \over 4 D t } \right)
&(8.14) \cr }
$$
where $D = kT / (2 M \gamma ) $. $g(y,t)$ obeys the diffusion equation
$$
{ \partial g \over \partial t } = D { \partial^2 g \over \partial
y^2 }
\eqno(8.15)
$$
{}From (8.14) or (8.15), one may compute all the moments of  $y$ at time $t$
in terms of their initial moments. One has, for example,
$$
\la y^4 \ra_t = 12 D^2 t^2 + 12 D t \la y^2 \ra_0  + \la y^4 \ra_0
\eqno(8.16)
$$
$$
\la y^5 \ra_t = 60 D^2 t^2 \la y \ra_0 + 20 D t \la y^3 \ra_0
+ \la y^5 \ra_0
\eqno(8.17)
$$
{}From the explicit expressions for the moments such as these, it is
straightforward to show that the leading order behaviour of all
moments of the form $ \la y^n \ra $
as $t \ria \infty $ are correctly reproduced by the Gaussian
distribution,
$$
g(y) = { 1 \over ( 4 \pi D t )^{\half} }
\ \exp \left( - { ( y - \la y \ra )^2 \over 4 D t } \right)
\eqno(8.18)
$$
It follows that for large times, the Wigner function approaches the
asymptotic form,
$$
W(p,q,t) = \exp \left( - {p^2 \over 2 M k T } \right)
\ \exp \left( - { M \gamma \over 2 k T t} \left( q - { p \over 2 M
\gamma } - \la q \ra_0 - { \la p \ra_0 \over 2 M \gamma } \right)^2
\right)
\eqno(8.19)
$$
Hence the distribution of $p$ is a Boltzmann distribution, as noted above,
and the $q$ distribution, obtained by integrating out $p$,
is peaked about the value
$ q = \la q \ra_0 + { \la p \ra_0 /  (2 M \gamma) } $, which is the
asymptotic value of $q$ under classical evolution, starting from the
initial values $\la q \ra_0$, $\la p \ra_0$.

The result (8.19) can, however, be rather misleading, and has limited
value as an approximation to the Wigner function at large times.
To understand this, write the Wigner function for large
times as
$$
W(p,q,t) = W_S (p,q,t) + W_1 (p,q,t) + ...
\eqno(8.20)
$$
where $W_S$ is the leading order approximation to the Wigner
function for large $t$ and $W_1$ is the next to leading order term.
Let us compare the free particle case considered here
with the case in which there
is a stationary solution. When a stationary solution exists, $W_S$
is the stationary solution, and hence is independent of time.
Furthermore, from Eq.(8.5), the next term $W_1$ is proportional to
$e^{- \lambda_1 t} $ where $ \lambda_1 > 0$ is the first eigenvalue.
It follows that all moments of (8.20) are given by their moments in
$W_S$ plus an exponential decaying term. Even if the moment of
$W_S$ vanishes, the correction given by $W_1$ goes to zero for large
$t$. All moments of (8.20) approach the stationary moments like $ e^{-
\lambda_1 t }$.

Now consider the free particle case. Here there is no stationary
solution, so $W_S$ is time dependent. The next correction $W_1$ does
not decay exponentially fast.
There are certain moments ({\it e.g.}, $\la ( y - \la y \ra )^3 \ra$)
that are zero for $W_S$ and
take their leading contribution from $W_1$.
Such moments {\it grow} with time, as may be seen from (8.16), (8.17),
so unlike the stationary solution case, they are
not well approximated by their moments in $W_S$.
Also, even for the moments which are well-approximated by the
moments in $W_S$ for large $t$, the rate of approach to the regime
in which that approximation is valid depends on the initial
conditions. It does not proceed at a universal rate.

Therefore, although Eq.(8.19) is the formal solution to the
Wigner function equation for large $t$, it is not very
useful. In practice it will be more useful to work directly with
the equations for the moments.

\head{\bf 9. Summary and Discussion}

We have studied the evolution of open quantum systems described
by the evolution equation (2.17). We were concerned with two particular
questions: generalized uncertainty relations for this class of
non-equilibrium systems and long-time limits. Our results may be summarized
as follows:

\item {\bf 1.} For any linear system whose evolution is described by the
propagator
(2.19)--(2.21), the uncertainty ${\cal U}$ and the Wigner function area
${\cal A}$ have the sharp
lower bounds (4.11), (4.18), respectively.
These represent the least possible amount of
noise the system must suffer after evolution for time $t$ in the presence of
an environment. These expressions are valid for all types of
environment ({\it i.e.}, for all choices of spectral density).
Also worthy of note is the particularly simple expression (4.17) of the
evolution of the Wigner function area $\A$.

\item {\bf 2.} For the particular case of the ohmic environment, the explicit
form of the lower bounds is given in Section V. These explicit
expressions give the comparative sizes of quantum and thermal
fluctuations, generalizing the work of Hu and Zhang [\cite{HuZ}].

\item {\bf 3.} For the linear systems considered here, these generalized
uncertainty relations achieve equality for Gaussian pure initial states
of the form (4.16). Such states are therefore the ones that suffer the
least amount of noise
under evolution in the presence of an environment,
substantiating the results of Zurek {\it et al.} [\cite{ZHP,PHZ,PaZ}].

\item {\bf 4.} The uncertainty is connected to the von Neumann entropy via
the relation (6.12), and the entropy is connected to the linear entropy by the
relation
(6.21).

\item {\bf 5.} For a harmonic oscillator in an ohmic environment,
all initial states tend towards a Gaussian Wigner function in the
long-time limit. It is the same as a thermal state in the high
temperature limit or the weak coupling limit.
For non-ohmic environments, the same result is
plausibly true for a reasonable class of spectral densities, but
this remains to be shown.

\item {\bf 6.} For general potentials such that $ \exp\left( - V_R(q) /kT
\right)$
is normalizable, the Wigner function tends towards a thermal state in the high
temperature limit, for all initial states. For the free particle,
the Wigner function tends to a Gaussian state,
although this is not a very useful expression because it does not
give a correct approximation to all the moments for large times.

The reason we were able to prove the results (1) and (4) with such ease was our
use of the Wigner function propagator, in terms of which the quantum evolution
takes a particularly transparent form.
We comment that the detailed methods used here could well be of use in related
calculations. For example, it might be possible to discuss decoherence of
arbitrary initial states using the Wigner function propagator derived in
Section III. These and other related questions will be pursued in
future publications.

\head {\bf Acknowledgements}

We would like to thank many of our colleagues for useful
conversations, including Bei-Lok Hu, Juan Pablo Paz and Wojciech Zurek.
C.A was supported by the Greek State Scholarship
Foundation.
J.J.H. was supported by a University Research Fellowship from the
Royal Society, and by the Isaac Newton Institute for Mathematical
Sciences, Cambridge, where part of this work was carried out.

\head{\bf Appendix}

In this appendix we give the explicit forms of the coefficients
$A,B,C$ and $\tilde K, \hat K, L, N$ appearing in the explicit expression
for the propagator, (2.17). The following results are taken from
Caldeira and Leggett [\cite{CaL}], and from Hu and Zhang [\cite{HuZ}], with
minor
elaborations and extensions.

We first give the forms of the coefficients for the harmonic
oscillator in the underdamped case, $\om_R > \gamma $.
Let $\om^2 = \om^2_R - \gamma^2 $. We work in the underdamped case,
$ \gamma < \omega_R$.

Then we have,
$$
\eqalignno{
\tilde K(t) &= - \half M \gamma + \half M  \omega
\cot \omega t,
&(A.1)\cr
\hat K(t) &= + \half M \gamma + \half M \omega \cot
\omega t,
&(A.2)\cr
L ( t) &= {M \omega e^{-\gamma t} \over 2
\sin  \omega t },
&(A.3)\cr
N ( t) &= {M \omega e^{\gamma t} \over 2
\sin \omega t }.
&(A.4)\cr }
$$
Also,
$$
A(t) = { M \gamma \over \pi} \int_0^{\infty} d \nu \ \exp
\left( - {\nu^2 \over \Lambda^2} \right)
\ \nu \ \coth \left( { \hbar \nu \over 2 k T } \right) \ A_{\nu} (t)
\eqno(A.5)
$$
where
$$
A_{\nu}(t) = { e^{-2 \gamma t} \over \sin^2 \om t }
\ \int_0^t d \tau \int_0^t ds
\ \sin \om \tau \ \cos \nu (\tau -s ) \ \sin \om s
\ e^{\gamma (\tau +s)}
\eqno(A.6)
$$
Similarly,
$$
B(t) = { M \gamma \over \pi} \int_0^{\infty} d \nu \ \exp
\left( - {\nu^2 \over \Lambda^2} \right)
\ \nu \ \coth \left( { \hbar \nu \over 2 k T } \right) \ B_{\nu} (t)
\eqno(A.7)
$$
where
$$
B_{\nu}(t) = { 2 e^{- \gamma t} \over \sin^2 \om t }
\ \int_0^t d \tau \int_0^t ds \
\sin \om \tau \ \cos \nu (\tau -s ) \ \sin \om (t-s)
\ e^{\gamma (\tau +s) }
\eqno(A.8)
$$
and
$$
C(t)= { M \gamma \over \pi} \int_0^{\infty} d \nu \ \exp
\left( - {\nu^2 \over \Lambda^2} \right)
\ \nu \ \coth \left( { \hbar \nu \over 2 k T } \right) \ C_{\nu} (t)
\eqno(A.9)
$$
where
$$
C_{\nu}(t) = { 1 \over \sin^2 \om t }
\ \int_0^t d \tau  \int_0^t ds \
\sin \om (t - \tau) \ \cos \nu (\tau -s ) \ \sin \om (t-s)
\ e^{\gamma (\tau +s) }
\eqno(A.10)
$$
We have included, for completeness, the explicit dependence on the cutoff in
the
expressions for $A$, $B$ and $C$, and this is sometimes required,
although we will generally work with the case
$ \Lambda \ria \infty $.  The integrals for $A_\nu$, $B_{\nu}$ and $C_{\nu}$
have been evaluated by Hu and Zhang [\cite{HuZ}],
and we will make heavy use of their results. The remaining integrals
over $\nu$ to yield $A$, $B$ and $C$ cannot be carried out in general,
but asymptotic expansions are possible in various regimes of interest, and
these we now give.

\subhead {\bf (A1). High-Temperature Limit}

In the much-studied high-temperature (Fokker-Planck) limit,
one has $\coth ( \hbar \nu / k T ) \approx k T / \hbar \nu $,
and the integrals (A.5)--(A.10) may be evaluated exactly for any $t$,
with the results,
$$
\eqalignno{
A & = { M k T \over 2 \hbar \sin^2 \om t}
\left[ 1 - e^{-2 \gamma t} - { 1 \over \gamma^2 + \om^2}
( \gamma^2 \cos 2 \om t + \om \gamma \sin 2 \om t )
- { \gamma \over \gamma^2 + \om^2} e^{-2 \gamma t} \right]
&(A.11) \cr
B & ={ M k T e^{\gamma t} \over  \hbar \sin^2 \om t}
\left[ - ( 1 - e^{-2 \gamma t} ) \cos \om t + { 1 \over \gamma^2 + \om^2}
\left( \gamma^2 ( 1 + e^{-2 \gamma t} ) \cos \om t \right. \right.
\cr &
\quad\quad\quad\quad\quad\quad\quad\quad\quad\quad\quad
\quad\quad\quad\quad\quad
\left. \left. + \om \gamma
( 1 - e^{-2 \gamma t} ) \sin \om t \right) \right]
&(A.12) \cr
C & = { M k T e^{ 2 \gamma t} \over 2 \hbar \sin^2 \om t}
\left[ { \om^2 \over \gamma^2 + \om^2 } - e^{-2 \gamma t}
+ { e^{-2 \gamma t} \over \gamma^2 + \om^2}
( \gamma^2 \cos \om t - \om \gamma \sin 2 \om t ) \right]
&(A.13) \cr }
$$
At short times, one has
$$
A = B = C = { 2 M \gamma k T t \over 3\hbar }
\eqno(A.14)
$$
and for long times,
$$
\eqalignno{
A & = { M k T \over 2 \hbar \sin^2 \om t}
\left[ 1 - { 1 \over \gamma^2 + \om^2}
( \gamma^2 \cos 2 \om t + \om \gamma \sin 2 \om t )  \right]
&(A.15) \cr
B & ={ M k T e^{\gamma t} \over  \hbar \sin^2 \om t}
\left[ - { \om^2  \over \gamma^2 + \om^2} \cos \om t
+ { \om \gamma \over \gamma^2 + \om^2 } \sin \om t \right]
&(A.16) \cr
C & = { M k T e^{ 2 \gamma t} \over 2 \hbar \sin^2 \om t}
{ \om^2 \over (\gamma^2 + \om^2) }
&(A.17) \cr }
$$

\subhead {\bf (A2). Weak Coupling Limit}

In the weak coupling regime, $ \gamma << \omega $, but for arbitrary
temperatures,
one has, from Hu and Zhang,
$$
\eqalignno{
A & = { M\om \over 4 \sin^2 \om t} \coth \left( { \hbar \om \over 2 kT}
\right) \left[ 1 - { \gamma \over \om} \sin 2 \om t - e^{ - 2\gamma t} \right]
&(A.18) \cr
B & = { M\om e^{\gamma t} \over 2 \sin^2 \om t} \coth\left( { \hbar \om \over
2kT} \right)
\left[ { \gamma \over \om} \sin \om t - \cos \om t + ( {\gamma \over \om}
\sin \om t + \cos \om t ) e^{-2 \gamma t} \right]
&(A.19) \cr
C & = {M \om e^{2 \gamma t} \over 4 \sin^2 \om t} \coth \left( {\hbar \om
\over 2 k T } \right) \left[ 1 - ( 1 + { \gamma \over \om } \sin 2 \om t )
e^{-2 \gamma t} \right]
&(A.20) \cr }
$$
where terms of order $ \gamma^2 / \om^2 $ have been neglected.
The long-time limits of these expressions are easily seen.

\subhead {\bf (A3). Long-Time Limit for Arbitrary Temperature and Coupling}

It is also possible to determine the exact form of the
long-time limits of $A$, $B$ and $C$,
without assuming high temperature or weak coupling. This is necessary in order
to give a completely general statement about the long-time limits of arbitrary
initial states, as in Section VII. From Hu and Zhang, they are
$$
\eqalignno{
A & = { M \over 2 \om_R^2 \sin^2 \om t}
\left[ ( \om^2 \cos^2 \om t - \gamma \om \sin 2 \om t)
(F_1 + { \gamma \over \om} F_2 )
\right.
\cr & \quad\quad\quad\quad\quad\quad\quad\quad\quad\quad
\left.
+ \om^2 \sin^2 \om t \left( ( 1 + { 2 \gamma^2 \over \om^2} ) F_1
- { \gamma \over \om} F_2 \right) \right]
&(A.21) \cr
B & = { M e^{\gamma t } \over \om_R^2 \sin^2 \om t} ( \gamma \om
\sin \om t -  \om^2 \cos \om t ) ( F_1 + {\gamma \over \om} F_2 )
&(A.22) \cr
C & = { M \om^2 e^{2 \gamma t} \over 2 \om_R^2 \sin^2 \om t}
( F_1 + { \gamma \over \om} F_2 )
&(A.23) \cr
}
$$
where
$$
\eqalignno{
F_1 & = {1 \over 2 \pi } \int_0^{\infty} d \nu \ \nu \coth\left(
{ \hbar \nu \over 2 kT} \right) \left( { \gamma \over \gamma^2 +
(\om + \nu)^2 } + { \gamma \over \gamma^2 + ( \om - \nu)^2 } \right)
&(A.24) \cr
F_2 & = {1 \over 2 \pi } \int_0^{\infty} d \nu \ \nu \coth\left(
{ \hbar \nu \over 2 kT} \right) \left( { \om + \nu \over \gamma^2 +
(\om + \nu)^2 } + { \om - \nu \over \gamma^2 + ( \om - \nu)^2 } \right)
&(A.25) \cr}
$$
The integrals $F_1$ and $F_2$ do not appear to be exactly soluble, but can be
evaluated
in the high temperature and weak couling regimes.
In the high temperature regime,
$$
F_1 \approx { k T \over \hbar}, \quad F_2 \approx 0
\eqno(A.26)
$$
and the results (A.11)--(A.13) are recovered. In the weak coupling regime,
$$
F_1 \approx \half \omega \coth\left( { \hbar \omega \over 2 k T } \right),
\quad F_2 \approx 0
\eqno(A.27)
$$
and the results (A.18)--(A.20) are recovered.

\subhead{\bf (A4). Very Short Time Limit with Finite Cutoff}

For finite cutoff, and in the very time limit, $ t << \Lambda^{-1} $,
and
at high temperature, $k T >> \hbar \omega $, one has
$$
A = 2 B = C = { M \gamma k T \Lambda \over 4 \pi \hbar } t^2
\eqno(A.28)
$$
and
$$
\eqalignno{
\tilde K &= { M \over 2 t } + { M \gamma \Lambda^3 \over 45 \pi } t^3
&(A.29) \cr
\hat K &= { M \over 2 t } +   { M \gamma \Lambda^3 \over 4 \pi} t^3
&(A.30) \cr
L &= { M \over 2 t} -       { 11 M \gamma \Lambda^3 \over 180 \pi} t^3
&(A.31) \cr
N &= { M \over 2 t } -         { M \gamma \Lambda^3 \over 180 \pi } t^3
&(A.32) \cr }
$$

\references

\def\pr{{\sl Phys. Rev.\ }}
\def\prl{{\sl Phys. Rev. Lett.\ }}
\def\prep{{\sl Phys. Rep.\ }}

\def\cmp{{\sl Comm. Math. Phys.\ }}

\def\pl{{\sl Phys. Lett.\ }}
\def\annp{{\sl Ann. Phys. (N.Y.)\ }}

\refis{Aga} G.S.Agarwal, \pr {\bf A3}, 828 (1971);
\pr {\bf A4}, 739 (1971).

\refis{Amb} V.Ambegaokar, {\sl Ber.Bunsenges.Phys.Chem.} {\bf 95},
400 (1991). Ambegaokar in turn cites a private communication from
P.Pechukas as the origin of the observation that the master equation
suffers from a problem with positivity. A modified master equation
that does preserve positivity was proposed by
L.Di\'osi, {\sl Europhys.Lett.} {\bf 22}, 1 (1993).

\refis{AnH} A.Anderson and J.J.Halliwell, \pr {\bf 48},
2753 (1993).

\refis{Ang} J.R.Anglin, ``Influence Functionals and Black Body
Radiation'', McGill preprint.

\refis{BaJ} N.Balazs and B.K.Jennings, \prep {\bf 104}, 347 (1984),
M.Hillery, R.F.O'Connell, M.O.Scully and E.P.Wigner, \prep {\bf
106}, 121 (1984);

\refis{BrK} V.B.Braginsky and F.Ya.Khalili, {\it Quantum
Measurement}
(Cambridge University Press, Cambridge, 1992).

\refis{Bru} T.Brun, \pr {\bf 47}, 3383 (1993).

\refis{CaL} A.O.Caldeira and A.J.Leggett, {\sl Physica} {\bf
121A}, 587 (1983).

\refis{Cov} T.M.Cover and J.A.Thomas,
{{\it Elements of Information Theory}} (Wiley, New York, 1991).

\refis{Dek} H.Dekker, \pr {\bf A16}, 2116 (1977); {\sl Phys.Rep.}
{\bf 80}, 1 (1991).

\refis{DKM} V.V.Dodonov, E.V.Kurmyshev and V.I.Man'ko, \pl
{\bf 79A}, 150 (1980).

\refis{DoH} H.F.Dowker and J.J.Halliwell, \pr {\bf D46}, 1580
(1992).

\refis{FeV} R.P.Feynman and F.L.Vernon, \annp {\bf 24}, 118 (1963).

\refis{GSI} H.Grabert, P.Schramm, G-L. Ingold, \prep {\bf 168},
115 (1988).

\refis{Hal1} J.J.Halliwell, \pr {\bf D46}, 1610 (1992).

\refis{Hal2} J.J.Halliwell, {\sl Phys.Rev.} {\bf D48}, 2739 (1993).

\refis{Hal3} J.J.Halliwell, {\sl Phys.Rev.D} {\bf D48}, 4785 (1993).

\refis{HuR} M.A.Huerta and H.S.Robertson, {\sl J.Stat.Phys.}
{\bf 1}, 393 (1969).

\refis{FLO} G.W.Ford, J.T.Lewis and R.F.O'Connell,
{\sl J.Stat.Phys.} {\bf 53}, 439 (1988).

\refis{HPZ} B.L.Hu, J.P.Paz and Y.Zhang, \pr {\bf D45}, 2843
(1992); \pr {\bf D47}, 1576 (1993).

\refis{HuZ} B.L.Hu and Y.Zhang, {\sl Mod.Phys.Lett.} {\bf A8},
3575 (1993).

\refis{Hus} K.Husimi, {\sl Proc.Phys.Math.Soc. Japan} {\bf 22},
264 (1940).

\refis{JoZ} E.Joos and H.D.Zeh, {\sl Z.Phys.} {\bf B59}, 223 (1985).

\refis{Lan} L.D.Landau and E.M.Lifshitz, {\it Statistical Physics}
(Pergammon, London, 1969), p.393.

\refis{Lie} E.H.Lieb, \cmp {\bf 62}, 35 (1978).

\refis{PHZ} J.P.Paz, S.Habib and W.Zurek, \pr {\bf D47}, 488 (1993).

\refis{PaZ} J.P.Paz and W.Zurek, \pr {\bf 48}, 2728 (1993).

\refis{Ris} H.Risken, {\it The Fokker-Planck Equations: Methods of
Solution and Applications}, Second Edition (Springer-Verlag, Berlin,1989).

\refis{Tat} V.I.Tatarskii, {\sl Sov.Phys.Usp} {\bf 26}, 311 (1983).

\refis{TeS} M.Tegmark and H.S.Shapiro, ``Decoherence Produces
Coherent States: An Explicit Proof for Harmonic Chains'',
preprint gr-qc/9402026 (1994).

\refis{UnZ} W.G.Unruh and W.H.Zurek, \pr {\bf D40}, 1071 (1989).

\refis{Weh} A.Wehrl, {\sl Rep.Math.Phys.} {\bf 16}, 353 (1979).

\refis{Zur} W.Zurek, {\sl Prof.Theor.Phys.} {\bf 89},
281 (1993);
and in, {\it Physical Origins of Time Asymmetry},
edited by J.Halliwell, J.Perez-Mercader and W.Zurek (Cambridge
University Press, Cambridge, 1994).

\refis {Zur2} W.H.Zurek, in {\it Frontiers of Non-Equilibrium
Statistical Mechanics}, edited by G.T.Moore and M.O.Scully (Plenum,
New York, 1986).

\refis{ZHP} W.Zurek, S.Habib and J.P.Paz, \prl {\bf 70},
1187 (1993).

\refis{ZuP} W.Zurek and J.P.Paz, ``Decoherence, Chaos and the
Second Law'', Los Alamos preprint (1993).

\endreferences

\end